\documentclass[12pt] {article}
\usepackage{amssymb} 
\newcommand{\vph}{{\varphi}}

\newcommand{\val}{{{\alpha}}}

\newcommand{\ovz}{{\overline z}}

\newcommand{\resection}[1]{\setcounter{equation}{0}\section{#1}}

\def\cc#1{\kern .7em\hfill #1 \hfill\kern .7em}

\newcommand{\nc}{\newcommand}
\newcommand{\tend}[1]{\raisebox{-0.2cm}{~\shortstack{ ${\rightarrow
}$ \\
${\vspace{-0.2cm} _{#1}}$   }}}
\nc{\beqa}{\begin{eqnarray}}
\nc{\eeqa}{\end{eqnarray}}
\def\agt{
\mathrel{\raise.3ex\hbox{$>$}\mkern-14mu\lower0.6ex\hbox{$\sim$}}
}
\def\alt{
\mathrel{\raise.3ex\hbox{$<$}\mkern-14mu\lower0.6ex\hbox{$\sim$}}
}

\begin {document}
\bibliographystyle{unsrt}    
\centerline{\Large \bf Expectation values of descendent fields}
\centerline{\Large \bf in the Bullough-Dodd model and}
\centerline{\Large \bf related perturbed conformal field theories}
\vskip 1cm

\centerline{\large P. Baseilhac\,\footnote{ \tt e-mail: pb18@york.ac.uk}
\ \ and\ \  M. Stanishkov\,\footnote{ \tt e-mail: marian@mail.apctp.org,
 On leave of absence from INRNE, Sofia, Bulgaria}}

\vspace{3mm} 

\small\normalsize
\vspace{0.5cm}

\centerline {\it $^{1}$ Department of Mathematics, University of York }

\vspace{2mm}

\centerline {\it Heslington, York YO105DD, United Kingdom}

\vspace{3mm}

\centerline {$^{2}$\it  Asia Pacific Center for Theoretical Physics,}

\vspace{2mm}

\centerline {\it Seoul, 130-012, Korea}

\vspace{1cm}

\begin{abstract}
The exact vacuum expectation values of the second level descendent fields\\
 $<(\partial\varphi)^2({\overline\partial}\varphi)^2e^{a\varphi}>$ \ in the
Bullough-Dodd model are calculated. By performing quantum group
restrictions, we obtain \ $<L_{-2}{\overline L}_{-2}{\Phi}_{lk}>$ \ in the
$\Phi_{12}$, $\Phi_{21}$ and $\Phi_{15}$ perturbed minimal CFTs. In
particular, the exact expectation value $<T{\overline
T}>$ is found to be proportional to the square of the bulk free energy.
\end{abstract}

\pagestyle{plain} 

\setcounter{page}{1}
\resection{Introduction}

In a 2-D integrable quantum field theory (QFT) which can be realized as
a conformal field theory (CFT) perturbed by some relevant operator, it
is well-known that any correlation function of local fields\ 
${\cal O}_{a}(x)$\ in the short-distance limit can be reduced down to one-point functions \
$<{\cal O}_{a'}(x)>$ by successive application of the operator product
expansion (OPE) \cite{2,3}. These vacuum expectation values (VEV)s
contain important information about the IR environment. Together with
the structure constants characterizing the UV limit of the QFT, they
provide the UV behaviour of the correlation
functions whereas the so-called form-factors characterize  their IR
behaviour. Since three years important progress has been made concerning the
evaluation of some VEVs in different integrable QFTs.  In ref. \cite{4}, an
explicit expression for the VEVs of the exponential fields in the sinh-Gordon
and sine-Gordon models was proposed.
In ref.
\cite{Fateev20,5} it was shown that this result can be obtained using the
``reflection amplitude'' \cite{6} of the Liouville field theory. This
method was also applied in the
so-called Bullough-Dodd model with real and imaginary coupling. In QFT
involving more fields,
 the VEVs for a two-parameter family of
integrable QFTs introduced and studied in \cite{9} gave rise to the VEVs of
local operators in parafermionic sine-Gordon models and in integrable
perturbed $SU(2)$ coset CFT \cite{9bis}. Also, the VEVs in simply-laced affine
Toda field theories are known for a long time \cite{norm} and
 the case of non-simply laced dual pairs was recently studied in
\cite{tba,pert}.

However, the higher-order corrections to the short-distance expansion of two-point
correlation functions involve the VEVs of the descendent fields. This question
was addressed in \cite{des}. There, the VEV of the descendent field \ \ 
$<(\partial\varphi)^2({\overline\partial}\varphi)^2e^{a\varphi}>$ in the
sinh-Gordon (ShG) and sine-Gordon (SG) - with $a\rightarrow i\alpha$ - models
was calculated as well as in its related perturbed CFT, i.e. $\Phi_{13}$
perturbation of minimal models. From this result, the next-order correction of
the two-point function in the scaling Lee-Yang model was computed \cite{3,5}.

The purpose of this paper is to calculate the VEV of the simplest
non-trivial descendent field in the Bullough-Dodd (BD) model which is
generally described by the following action in the Euclidean space :
\beqa
{\cal A}_{BD} = \int d^2x \big[\frac{1}{16\pi}(\partial_\nu\varphi)^2 +
\mu e^{b\varphi} + \mu' e^{-\frac{b}{2}\varphi}\big].\label{actionBD}
\eeqa
Here, the parameters $\mu$ and $\mu'$ are introduced, as the two operators do
not renormalize in the same way, on the contrary to any simply-laced affine
Toda field theory. This model has attracted over the years a certain
interest, in particular in connection with perturbed minimal models :  
 $c<1$ minimal CFT perturbed  by the operators
$\Phi_{12}$, $\Phi_{21}$ or $\Phi_{15}$ can be obtained by a quantum group (QG) restriction
of imaginary Bullough-Dodd model \cite{smir,cost,tak,tak2,5} with
special values of the coupling. We will use this property to
deduce the VEV  \ $<L_{-2}{\overline L}_{-2}{\Phi}_{lk}>$ in the
following perturbed minimal models :
\beqa
{\cal{A}} &=& {\cal{M}}_{p/p'} + {\lambda} \int d^2x
\Phi_{12}\ ,\label{action}\\
{\hat{\cal{A}}} &=& {\cal{M}}_{p/p'} + 
{\hat\lambda} \int d^2x \Phi_{21}\ \label{actiontilde}\\
\mbox{or}\ \ \ \ \ \ {\tilde{\cal{A}}} &=& {\cal{M}}_{p/p'} + 
{\tilde\lambda} \int d^2x \Phi_{15}\ ,\label{actionhat}
\eeqa
where  we denote respectively $\Phi_{12}$, $\Phi_{21}$ and $\Phi_{15}$ 
as specific primary operators of the unperturbed minimal model
${\cal{M}}_{p/p'}$ and the
parameters $\lambda$, $\hat\lambda$ and $\tilde\lambda$ 
\ characterize the strength of the perturbation.

This paper is organized as follows.
In the next section we introduce the notations and write the
short-distance expansion of the two-point correlation function which
involves the VEV of the descendent field  
 \ $<(\partial\varphi)^2({\overline\partial}\varphi)^2e^{a\varphi}>$ in
the BD model associated with the action (\ref{actionBD}). Using the
method based on the ``reflection relations'' \cite{6} we find a conjecture for this last
quantity in Section 3. Whereas it exists an ambiguity for the solution of these
functional equations, we choose the ``minimal'' one which is compatible
with the ``resonance conditions'' (see ref. \cite{des} for details). 
In Section 4 we compare the semi-classical limit of the short distance
expansion of the two-point function with the semi-classical approximation based
on the action (\ref{actionBD}).  In  Section 5 we deduce the VEV 
$<L_{-2}{\overline L}_{-2}{\Phi}_{lk}>$ in the models (\ref{action}),
(\ref{actiontilde}) and (\ref{actionhat}).  Concluding remarks follow in the last
section. 
\resection{Short-distance expansion of the two-point function in the BD model}

Similarly to the ShG model \cite{des}, the BD model can be regarded as a
relevant perturbation of a Gaussian CFT. In this free field theory, the field
is normalized such that:
\beqa
<\varphi_(z,{\ovz})\varphi(0,0)>_{Gauss}=-2\log(z{\ovz}).
\eeqa
and we have the classical equation of motion :
\beqa
\partial {\overline \partial}\varphi = 0.\label{motion}
\eeqa
Instead of considering the action (\ref{actionBD}) we turn directly to the
case of an imaginary coupling constant which is the most interesting for
our purpose in Section 5. The perturbation is then relevant if \
$0<\beta^2<1$. Although the model (\ref{actionBD}) 
for real coupling is very different from the one with imaginary coupling
in its physical content (this latter model contains solitons
 and breathers), there are good
reasons to believe that the expectation values obtained in the real
coupling case provide also the expectation values for the imaginary coupling.
The calculation of the VEVs in both cases ($b$ real or imaginary) within
 the standard perturbation theory agree through the identification
 $b=i\beta$ \cite{5}. With this substitution in (\ref{actionBD}), 
the general short distance OPE for two arbitrary primary fields 
$e^{i\alpha_1\varphi}(x)$ and $e^{i\alpha_2\varphi}(y)$ takes the form :
\beqa
e^{i\alpha_1\varphi}(x)e^{i\alpha_2\varphi}(y) &=& 
\sum_{n=0}^{\infty}\big\{
C_{\alpha_1\alpha_2}^{n,0}(r)e^{i(\alpha+n\beta)\vph}(y)+ \ ... \big\}\nonumber\\
&+&\sum_{n=1}^{\infty}\big\{{C'}_{\alpha_1\alpha_2}^{\ n,0}(r)e^{i(\alpha-\frac{n\beta}{2})\vph}(y)+ \ ...\big\}\nonumber \\
&+&
\sum_{n=1}^{\infty}\big\{D_{\alpha_1\alpha_2}^{n,0}(r)e^{i(\alpha+(n-\frac{1}{2})\beta)\vph}(y)+
 \ ...\big\}\label{ope}
\eeqa
where $\alpha=\alpha_1+\alpha_2$, \ $r=|x-y|$ and  the dots in each term
stand for the contributions of the descendents of each field. The
different coefficients in eq. (\ref{ope}) are computable within the
conformal perturbation theory (CPT) \cite{3,dots}. We obtain :
\beqa
C^{n,0}_{\alpha_1\alpha_2}(r)&=&
 {\mu}^n r^{4\alpha_1\alpha_2+4n\beta(\alpha_1+\alpha_2)+2n(1-\beta^2)+2n^2\beta^2}
f^{n,0}_{\alpha_1\alpha_2}\big(\mu(\mu')^2
r^{6-3\beta^2}\big)\label{coef};\\
{C'}_{\alpha_1\alpha_2}^{\ n,0}(r)&=&
{\mu'}^n
r^{4\alpha_1\alpha_2-2n\beta(\alpha_1+\alpha_2)+2n(1-\frac{\beta^2}{4})+
\frac{n^2\beta^2}{2}}
{f'}^{n,0}_{\alpha_1\alpha_2}\big(\mu(\mu')^2 r^{6-3\beta^2}\big);\nonumber\\
D_{\alpha_1\alpha_2}^{\ n,0}(r)&=&
{\mu'}{\mu}^n
r^{4\alpha_1\alpha_2+4(n-\frac{1}{2})\beta(\alpha_1+\alpha_2)+2n(1-2\beta^2)+2+2n^2\beta^2}
{g}^{n,0}_{\alpha_1\alpha_2}\big(\mu(\mu')^2 r^{6-3\beta^2}\big)\nonumber
\eeqa
where any function $h\in\{f,{f'},g\}$ admits a power series expansion :
\beqa
h^{n,0}_{\alpha_1\alpha_2}(t)=\sum_{k=0}^{\infty}h^{n,0}_{k}(\alpha_1,\alpha_2)t^k.\label{serie}
\eeqa
Each coeffficient in (\ref{coef}) is expressed in terms of Coulomb type
integrals. The corresponding leading terms are respectively given by :
\beqa
f^{n,0}_{0}(\alpha_1,\alpha_2)&=&j_n(\val_1\beta,\val_2\beta,\beta^2)\ \ \ \ \mbox{for}\ \ \ n\neq
0\ ;\label{func}\\
{f'}^{n,0}_{0}(\alpha_1,\alpha_2)&=&j_n(-\frac{\val_1\beta}{2},-\frac{\val_2\beta}{2},
\frac{\beta^2}{4})\ ;\nonumber\\
{g}^{n,0}_{0}(\alpha_1,\alpha_2)&=&
{\cal F}_{n,1}(\val_1\beta,\val_2\beta,\beta^2)\nonumber
\eeqa
where we introduced the integrals :
\beqa
j_{n}(a,b,\rho)&=&\frac{1}{n!}\int\prod_{k=1}^{n}d^2 x_k \prod_{k=1}^{n}
 |x_k|^{4a} |1-x_k|^{4b} \prod_{k<p}^{n} |x_k-x_p|^{4\rho}\ ;\\ 
{\cal F}_{n,m}(a,b,\rho)&=&\frac{1}{n!m!}\int\prod_{k=1}^{n}d^2 x_k\int
\prod_{l=1}^{m} d^2y_l \prod_{k=1}^{n}
 |x_k|^{4a} |1-x_k|^{4b} \prod_{k<p}^{n} |x_k-x_p|^{4\rho}\nonumber\\
&&\ \ \ \ \ \ \ \ \ \ \times \ 
 \prod_{l=1}^{m}|y_l|^{-2a} |1-y_l|^{-2b} \prod_{l<q}^{m}
|y_l-y_q|^{\rho} \prod_{k,l}^{n,m}|x_k-y_l|^{-2\rho}\ .\nonumber
\eeqa
Notice that $f^{0,0}_{0}(\alpha_1,\alpha_2) =1$ and that the first subleading term of the
coefficient $C^{0,0}_{\val_1\val_2}$ is :
\beqa
f^{0,0}_{1}(\alpha_1,\alpha_2)={\cal F}_{1,2}(\val_1\beta,\val_2\beta,\beta^2)\ .
\eeqa
The integrals $j_n(a,b,\rho)$ have been evaluated explicitly in 
\cite{dots}  with the result :
\beqa
j_{n}(a,b,\rho)&=&\pi^n\big(\gamma(\rho)\big)^{-n}\prod_{k=1}^n \ \gamma(k\rho)
\times \label{jn}\\
&&\ \ \ \prod_{k=0}^{n-1}\gamma(1+2a+k\rho) \gamma(1+2b+k\rho) \gamma(-1-2a-2b-(n-1+k)\rho)\nonumber 
\eeqa
where the usual notation $\gamma(x)=\Gamma(x)/\Gamma(1-x)$ is used.

 As we already said in the introduction, the
next sub-leading terms in (\ref{ope}) involve the descendent
fields. There are four independent second-level descendent fields in BD :
\beqa
&&(\partial\vph)^2({\overline\partial}\vph)^2e^{i\alpha\vph}\ ;\ \ \ \ \ 
(\partial\vph)^2({\overline\partial}^2\vph)e^{i\alpha\vph}\ ;\label{4fields}\\
&&(\partial^2\vph)({\overline\partial}\vph)^2e^{i\alpha\vph}\ ;\ \ \ \ \ 
(\partial^2\vph)({\overline\partial}^2\vph)e^{i\alpha\vph}.\nonumber
\eeqa
Similarly to the SG (or ShG) case, using (\ref{motion}) it is easy to
show that linear combinations of these descendent fields can be written in
terms of total derivatives of local fields (we refer the reader to
\cite{des} for details about these relations). As a result,
 the VEVs of the composite fields (\ref{4fields}) can all be expressed in terms of 
a single VEV, say :
\beqa
<(\partial\vph)^2({\overline\partial}\vph)^2e^{i\alpha\vph}>_{BD}.\label{twovev}
\eeqa

Let us make an important observation. The second sub-leading terms in the OPE
(\ref{ope}) appear to be the third order descendents of the primary fields.
Analogously to the previous discussion linear combinations of them can be
expressed in terms of total derivatives of some local fields. As before, all the
corresponding VEVs can be expressed through
$<(\partial\varphi)^3(\bar\partial\varphi)^3e^{i\alpha\varphi}>$. Unlike the SG
case, it is non-vanishing due to the absence of a conserved charge of spin 3 in
the BD model. We will not enter in details about this VEV since its computation
is not our purpose in this paper.

One can now write the short-distance expansion for the
two-point function :
\beqa
{\cal G}_{\val_1\val_2}(r)=<e^{i\val_1\vph}(x)e^{i\val_2\vph}(y)>_{BD}\ \ \ \ \mbox{with} \ \ r=|x-y|
\eeqa
by taking the expectation value of the r.h.s. of the OPE (\ref{ope}) in the BD
model with imaginary coupling. Due to the previous discussion,
  the first non-vanishing contribution of the VEVs of lowest descendent
fields in the r.h.s. of the VEV of (\ref{ope}) correspond to the following terms :
\beqa
&&C^{n,2}_{\val_1\val_2}(r)
<(\partial\vph)^2({\overline\partial}\vph)^2
e^{i(\alpha+n\beta)\vph}>_{BD}\ ; \\ 
&&{C'}^{\ n,2}_{\val_1\val_2}(r)
<(\partial\vph)^2({\overline\partial}\vph)^2
e^{i(\alpha-\frac{n\beta}{2})\vph}>_{BD}\ ; \nonumber\\ 
&&D^{n,2}_{\val_1\val_2}(r)
<(\partial\vph)^2({\overline\partial}\vph)^2
e^{i(\alpha+(n-\frac{1}{2})\beta)\vph}>_{BD}\ ,\nonumber 
\eeqa
respectively. These coefficients also admit expansion similar to eqs.
 (\ref{coef}), (\ref{serie}) and (\ref{func}). In particular we have :
\beqa
C^{0,2}_{\val_1,\val_2}(r)=\frac{(\val_1\val_2)^2}{4}r^{4\alpha_1\alpha_2+4}\Big( \ 1\ + \
O\big(\mu(\mu')^2r^{6-3\beta^2}\big)\Big)\ .
\eeqa
Finally, the short-distance ($r\rightarrow 0$)
expansion of the two-point correlation function in
the BD model with imaginary coupling writes :
\beqa
&&{\cal G}_{\val_1\val_2}(r)= {\cal G}_{\val_1+\val_2} r^{4\val_1\val_2} 
\Big\{ 1 + {\cal F}_{1,2}(\val_1\beta,\val_2\beta,\beta^2)
\mu(\mu')^2r^{6-3\beta^2}  +\frac{(\val_1\val_2)^2}{4} {\cal H}(\val_1+\val_2)r^4
\nonumber \\
&& \ \ \ \ \ \ \qquad \ \  \ \qquad \qquad \ - \frac{\alpha_1^2\alpha_2^2(\alpha_1\!-\!\alpha_2)^2}{144}{\cal
K}(\alpha_1+\alpha_2)r^{6} 
 + O\big(\mu^2(\mu')^4r^{12-6\beta^2}\big) \Big\}\nonumber\\
&& + \sum_{n=1}^{\infty} {\mu}^n
r^{4\alpha_1\alpha_2+4n\beta(\alpha_1+\alpha_2)+2n(1-\beta^2)+2n^2\beta^2}
j_n(\val_1\beta,\val_2\beta,\beta^2)\nonumber \\
&& \ \ \ \ \ \ \ \ \  \ \ \ \ \ \ \  \ \ \ \ \  \ \ \ \ \ \ \ \ \ \ \ \
\ \ \ \ \ \ \  \ \ \ \ \ \ \   \times \ \ {\cal G}_{\val_1+\val_2+n\beta}
\ \Big\{\ 1+\ O\big(\mu(\mu')^2r^{6-3\beta^2}\big)\Big\}\nonumber \\
&& + \sum_{n=1}^{\infty} {\mu'}^n
r^{4\alpha_1\alpha_2-2n\beta(\alpha_1+\alpha_2)+2n(1-\frac{\beta^2}{4})+\frac{n^2\beta^2}{2}}
j_n(-\frac{\val_1\beta}{2},-\frac{\val_2\beta}{2},\frac{\beta^2}{4})\label{twop} \\
&& \ \ \ \ \ \ \ \ \  \ \ \ \ \ \ \  \ \ \ \ \  \ \ \ \ \ \ \ \ \ \ \ \
\ \ \ \ \ \ \  \ \ \ \ \ \ \   \times \ \  {\cal G}_{\val_1+\val_2-\frac{n\beta}{2}}
\ \Big\{\ 1+\ O\big(\mu(\mu')^2r^{6-3\beta^2}\big) \Big\}\nonumber \\
&& + \sum_{n=1}^{\infty} {\mu}^n{\mu'}
r^{4\alpha_1\alpha_2+4(n-\frac{1}{2})\beta(\alpha_1+\alpha_2)+2n(1-2\beta^2)+2+2n^2\beta^2}
{\cal F}_{n,1}(\val_1\beta,\val_2\beta,\beta^2)\nonumber \\
&& \ \ \ \ \ \ \ \ \  \ \ \ \ \ \ \  \ \ \ \ \  \ \ \ \ \ \ \ \ \ \ \ \
\ \ \ \ \ \ \  \ \ \ \ \ \ \   \times \ \  {\cal G}_{\val_1+\val_2+(n-\frac{1}{2})\beta}
\ \Big\{\ 1+\ O\big(\mu(\mu')^2r^{6-3\beta^2}\big) \Big\}\nonumber
\eeqa
where we defined ${\cal H}(\alpha)$ and ${\cal K}(\alpha)$
by the ratios :
\beqa
&&{\cal H}(\alpha)=
\frac{<(\partial\vph)^2({\overline\partial}\vph)^2e^{i\alpha\vph}>_{BD}} 
{<e^{i\alpha\vph}>_{BD}}\ ,\label{H}\\
&&{\cal K}(\alpha)=
\frac{<(\partial\vph)^3({\overline\partial}\vph)^3e^{i\alpha\vph}>_{BD}}
{<e^{i\alpha\vph}>_{BD}}\label{Ka} 
\eeqa 
and\ ${\cal G}_\val=<e^{i\alpha\vph}>_{BD}$ is the VEV of the exponential
field in the BD model. A closed analytic expression for this latter VEV
has been proposed in ref. \cite{5} :
\beqa
<\!e^{i\alpha\vph}\!>_{BD}\!\!\!&=&\!\!\Big[\frac{\mu'}{\mu}\frac{2^{\frac{-\beta^2}{2}}
\Gamma(1+\beta^2)\Gamma(1-\frac{\beta^2}{4})}{\Gamma(1-\beta^2)
\Gamma(1+\frac{\beta^2}{4})}\Big]^{\frac{2\alpha}{3\beta}}
\Big[\frac{m\Gamma(1-\frac{\beta^2}{6-3\beta^2})\Gamma(\frac{2}{6-3\beta^2})}
{2^{\frac{2}{3}}\sqrt{3}\Gamma(\frac{1}{3})}\Big]^{-\alpha\beta+2\alpha^2}
\times\nonumber \\
\!\!\!&&\!\!\!\exp\Big[\int^{+\infty}_0 \frac{dt}{t}
\Big(\frac{\sinh((2-\beta^2)t)\Psi(t,\alpha)}{\sinh(3(2-\beta^2)t)\sinh(2t)\sinh(\beta^2t)}-2\alpha^2e^{-2t}\Big)\Big]
\label{VEVBD}
\eeqa
where 
\beqa
\Psi(t,\alpha)&=&-\sinh(2\alpha\beta t)\big(\sinh((4-\beta^2-2\alpha\beta)t)
-\sinh((2-2\beta^2+2\alpha\beta)t)+\nonumber\\
&&\sinh((2-\beta^2-2\alpha\beta)t)-\sinh((2-\beta^2+2\alpha\beta)t)-\sinh((2+\beta^2-2\alpha\beta)t)\big).\nonumber
\eeqa
Its integral representation is well defined if :
\beqa
-\frac{1}{2\beta}\ <\ {\mathfrak R}e(\alpha)\ <\ \frac{1}{\beta}\label{cond}
\eeqa
and obtained by analytic continuation outside this domain. 

It is then straightforward to obtain the result associated with the action
(\ref{actionBD}) i.e. for {\it real} values of the coupling constant $b$ which
follows from the obvious substitutions :
\beqa
&&\beta\rightarrow -ib ;\ \ \ \val_1\rightarrow -ia_1\ ;\ \ \ \val_2\rightarrow
-ia_2\ ;\label{substit}\\
&&\mu\rightarrow -\mu\ ;\ \ \ \ \mu'\rightarrow -\mu' \ .\nonumber
\eeqa

In the (Gaussian) free field theory, the composite fields
(\ref{4fields}) are spinless with scale dimension :
\beqa
D\equiv \Delta+{\overline \Delta}=2\alpha^2+4.\label{dim}
\eeqa
For generic value of the coupling $\beta$ some divergences arise in
the VEVs of the fields
(\ref{4fields}) due to the perturbation in (\ref{actionBD}) with
imaginary coupling. They are generally cancelled if we add specific
counterterms which contain spinless local fields with cutt-off
dependent coefficients. For $0<\beta^2<1$ the perturbation becomes
relevant and a finite number of lower scale dimension couterterms are
then sufficient. However, this procedure is regularization scheme dependent,
i.e. one can always add finite counterterms. For generic values of
$\alpha$ this ambiguity in the definition of the renormalized expression
for the fields (\ref{4fields}) can be eliminated by fixing their scale
dimensions to be (\ref{dim}). It exists however a set of values of
$\alpha$ for which the ambiguity still remains, but here we will not
consider these isolated cases. In the BD model
with imaginary coupling, this situation arises if two fields, say 
${\cal O}_{\val}$ and  ${\cal O}_{\val'}$, satisfy the resonance condition :
\beqa
D_\val=D_{\val'} + 2n(1-\beta^2) + 2m(1-\frac{\beta^2}{4})\ \ \ \ \mbox{with}\
\ \ \ (n,m)\in {\mathbb N}
\eeqa
associated with the ambiguity :
\beqa
{\cal O}_{\val}\longrightarrow{\cal O}_{\val} + {\mu}^{n}{\mu'}^{m}{\cal O}_{\val'}.
\eeqa
In this specific case one says that the renormalized
field ${\cal O}_\val$ has an $(n|m)$-th {\it resonance}\,\footnote{The same
situation arise in any more general
affine Toda theory.} \cite{des} with the field ${\cal O}_{\val'}$. 
Due to the condition (\ref{cond}) and using (\ref{dim}) we find
immediatly that a resonance can appear between the descendent 
field $(\partial\varphi)^2(\bar\partial\varphi)^2e^{i\alpha\varphi}$ and the
following primary fields : 
\beqa
(i) \ \ \ \ &&e^{i(\val+\beta)\vph}\ \ \ \ \mbox{i.e.}\ \  (n|m)=(1|0) \ \ \
\mbox{for}\ \ \ \alpha=\frac{1}{2\beta}\ ;\label{nm}\\
(ii) \ \ \ \ &&e^{i(\val+2\beta)\vph}\ \ \ \mbox{i.e.}\ \  (n|m)=(2|0) \ \ \
\mbox{for}\ \ \ \alpha=-\frac{\beta}{2}\ ;\nonumber\\
(iii) \ \ \ \ &&e^{i(\val-\beta)\vph}\ \ \ \ \mbox{i.e.}\ \  (n|m)=(0|2) \ \ \
\mbox{for}\ \ \ \ \alpha=\frac{\beta}{4}\ ;\nonumber\\
(iv) \ \ \ \ &&e^{i(\val+\frac{\beta}{2})\vph}\ \ \ \ \mbox{i.e.}\ \  (n|m)=(1|1) \ \ \
\mbox{for}\ \ \ \ \alpha=\beta\ .\nonumber
\eeqa
If we now look at the expression (\ref{twop}), we notice that the
contribution (\ref{H}), brought by the second level descendent field, 
and that of any of the
exponential fields in  $(i)$, $(ii)$, $(iii)$ and $(iv)$, have
the same power behaviour in $r$ 
($r^{4\val_1\val_2 + 4}$) at short-distance for
the corresponding values of $\alpha$ in  (\ref{nm}). The integrals which
appear in these contributions and their corresponding poles are,
respectively :
\beqa
&&j_1(\val_1\beta,\val_2\beta,\beta^2)\ \ \ \ \ \ \ \ \ \ \ \ \mbox{with the pole}\ \ \ 
\alpha=\frac{1}{2\beta}\ ;\label{jpole}\\
&&j_2(\val_1\beta,\val_2\beta,\beta^2)\ \ \ \ \ \ \ \ \ \ \ \ \mbox{with
the pole}\ \ \ 
\alpha=-\frac{\beta}{2}\ ;\nonumber\\
&&j_2(-\frac{\val_1\beta}{2},-\frac{\val_2\beta}{2},\frac{\beta^2}{4})\ \ \ \ \
\mbox{with the pole}\ \ \ 
\alpha=\frac{\beta}{4}\ ;\nonumber\\
&&{\cal F}_{1,1}(\val_1\beta,\val_2\beta,\beta^2)\ \ \ \ \ \ \ \ \
 \mbox{with the pole}\ \ \ 
\alpha=\beta\ .\nonumber
\eeqa
By analogy with the SG (or ShG) model, one expects that the VEV
(\ref{H}) (and similarly for the real coupling case) exhibits, at least,
the same poles in order that the divergent contributions compensate each
other. This last requirement leads for instance to the relations :
\beqa
(i')\ \ \ && {\cal R}es_{\alpha=\frac{1}{2\beta}}{\cal H}(\val) =
8\pi\beta^3\mu\frac{{\cal G}_{\val+\beta}}{{\cal
G}_\val}|_{\val=\frac{1}{2\beta}}\ ;\label{residu}\\
(ii')\ \ \ && {\cal R}es_{\alpha=-\frac{\beta}{2}}{\cal H}(\val) =
-32\pi^2\beta^3\mu^2\frac{\gamma(2\beta^2)}{\gamma(\beta^2)}\gamma(-1-\beta^2)\frac{{\cal G}_{\val+2\beta}}{{\cal
G}_\val}|_{\val=-\frac{\beta}{2}}\ ;\nonumber\\
(iii')\ \ \ && {\cal R}es_{\alpha=\frac{\beta}{4}}{\cal H}(\val) =
4\pi^2\beta^3{\mu'}^2\frac{\gamma(\beta^2/2)}{\gamma(\beta^2/4)}\gamma(-1-\beta^2/4)\frac{{\cal G}_{\val-\beta}}{{\cal
G}_\val}|_{\val=\frac{\beta}{4}}\ ;\nonumber\\
(iv')\ \ \ && {\cal R}es_{\alpha=\beta}{\cal H}(\val) =
-\frac{4}{(\alpha_1\alpha_2)^2}\mu{\mu'}\frac{{\cal
G}_{\val+\frac{\beta}{2}}}{{\cal G}_\val}|_{\val=\beta}{\cal
R}es_{\alpha=\beta}\
{\cal F}_{1,1}(\alpha_1\beta,\alpha_2\beta,\beta^2)\ .\nonumber
\eeqa
These last conditions will be used in the next section to
fix the normalization of the VEV (\ref{H}). Let us now turn to the evaluation of
(\ref{H}) which plays an important role in the two-point function (\ref{twop}).
\resection{Reflection relations and descendent fields}
The BD model (\ref{actionBD}) can be regarded as two different
perturbations of the Liouville field theory \cite{5}. First, one can consider
the Liouville action :
\beqa
{\cal A}_{L}^{(1)} = \int d^2x \big[\frac{1}{16\pi}(\partial_\nu\varphi)^2 +
\mu e^{b\varphi}\big].\label{liouv}
\eeqa
The perturbation is then identified with $e^{-\frac{b}{2}\vph}$.
Alternatively, we can take :
\beqa
{\cal A}_{L}^{(2)} = \int d^2x \big[\frac{1}{16\pi}(\partial_\nu\varphi)^2 +
\mu' e^{-\frac{b}{2}\varphi}\big] \label{liouvtwo}
\eeqa
as the initial action and consider $e^{b\varphi}$ as a perturbation.
Using the first picture, the holomorphic stress-energy tensor :
\beqa
T(z)=-\frac{1}{4}(\partial\vph)^2 +
\frac{Q}{2}\partial^2\vph\label{str}
\eeqa
ensures the local conformal invariance of the Liouville field theory
(\ref{liouv}) and similarly for the anti-holomorphic part. The
exponential fields $e^{a\vph}$ are spinless primary fields with
conformal dimension :
\beqa
\Delta=a(Q-a).
\eeqa
The property of reflection relations which relates operators with the
same quantum numbers is a characteristic of the CFT. Using the
CPT framework, one expects that similar relations are also satisfied in
the perturbed case (\ref{actionBD}). With the change $b\rightarrow -b/2$
in (\ref{str}) and using the
second picture (\ref{liouvtwo}), one assumes that the VEV of the exponential field
$<e^{a\vph}>_{BD}$ satisfies simultaneously the following 
two functional equations \cite{5} : 
\beqa
&&<e^{a\vph}>_{BD}\ \ \ = \ \ R(a)<e^{(Q-a)\vph}>_{BD}\ ;\label{sys}\\
&&<e^{-a\vph}>_{BD}\ =\ \ R'(a)<e^{(-Q'+a)\vph}>_{BD}\nonumber
\eeqa
with 
\beqa
Q=\frac{1}{b}+b\ \ \ \ \ \mbox{and}\ \ \ \ \ Q'=\frac{2}{b}+\frac{b}{2}.
\eeqa
The functions $R(a)$, $R'(a)$ are called ``reflection amplitudes''.
An exact expression for $R(a)$ was presented in 
\cite{6}. $R'(a)$ is obtained from $R(a)$ by the substitutions
$b\rightarrow\frac{b}{2}$ and $\mu\rightarrow\mu'$. Under certain
assumptions about the analytic properties of the VEV, the system (\ref{sys})
was solved and the VEV for these exponential
fields was derived in \cite{5}.

Let us denote the descendent fields :
\beqa
L_{[n]}{\overline L}_{[m]}
e^{a\vph}\equiv L_{-n_1}...L_{-n_1}{\overline L}_{-m_1}...{\overline L}_{-m_K} e^{a\vph}\label{Lnm}
\eeqa
where $[n]=[-n_1,...,-n_N]$ \ and \ $[m]=[-m_1,...,-m_K]$ are arbitrary strings
and $L_n$, ${\overline L}_n$ are the standard Virasoro generators :
\beqa
T(z)=\sum_{n\in{\mathbb Z}}L_n z^{-n-2} \ \ \ \
 \mbox{and} \ \ \ \  {\overline T}({\overline z})=\sum_{n\in{\mathbb Z}}{\overline L}_n
{\overline z}^{-n-2}
\ .
\eeqa
The descendent fields (\ref{Lnm}) and the ones obtained after the
reflection $a\rightarrow Q-a$ possess the same quantum numbers.
Consequently, using the arguments of \cite{5,des}  based on the CPT
framework, one also expects that their VEVs in the perturbed theory
(\ref{actionBD}) satisfy the following ``reflection relation'' :
\beqa
<L_{[n]}{\overline L}_{[m]}e^{a\vph}>_{BD}=R(a)<L_{[n]}
{\overline L}_{[m]}e^{(Q-a)\vph}>_{BD}.\label{refLnm}
\eeqa
However, it is more convenient to use the basis : 
\beqa
(\partial^{n_1}\vph)...(\partial^{n_N}\vph)
({\overline\partial}^{m_1}\vph)...({\overline\partial}^{m_K}\vph)e^{a\vph}.\label{basis}
\eeqa
The main reason is that in (\ref{refLnm}) the components $L_n$, ${\overline
L}_n$ of the modified stress-tensor depend on $a$. Using (\ref{motion}) one can always
express (\ref{Lnm}) in the
basis (\ref{basis}). For our purpose we will need the relation \cite{6} :
\beqa
 L_{-2}{\overline L}_{-2}e^{a\vph}=\big[-\frac{1}{4}(\partial\vph)^2
+(\frac{Q}{2}+a)\partial^2\vph\big]\big[-\frac{1}{4}({\overline\partial}\vph)^2
+(\frac{Q}{2}+a){\overline\partial}^2\vph\big]e^{a\vph}\ .
\eeqa
Furthermore, using (\ref{motion}) it implies :
\beqa
<L_{-2}{\overline
L}_{-2}e^{a\vph}>_{BD}=\frac{1}{16}\big(1+2a(Q+2a)\big)^2
<(\partial\vph)^2({\overline\partial}\vph)^2e^{a\vph}>_{BD}\label{L2L2}
\eeqa
which leads to the following reflection relation :
\beqa
&&\big(1+2a(Q+2a)\big)^2
<(\partial\vph)^2({\overline\partial}\vph)^2e^{a\vph}>_{BD}=\\
&&\ \ \ \ \ \ \ \ \ \ \ \ \ \ \ \ \ \ \ \ \ \ \  \ \ \big(1+2(Q-a)(3Q-2a)\big)^2
<(\partial\vph)^2({\overline\partial}\vph)^2e^{(Q-a)\vph}>_{BD}\nonumber
\eeqa
One can also consider the second picture (\ref{liouvtwo}) where the Liouville
theory has coupling $-\frac{b}{2}$ instead of $b$ and is perturbed by
 $e^{b\vph}$. If we define the analytic continuation of (\ref{H}) :
\beqa
H(a)=\frac{<(\partial\vph)^2({\overline\partial}\vph)^2e^{a\vph}>_{BD}}
{<e^{a\vph}>_{BD}},\label{Hreal}
\eeqa
then the two different pictures provide us the following two functional
relations :  %
\beqa
H(a)&=&\Big[\frac{(2b+3/b-2a)(3b+2/b-2a)}{(b+2a)(1/b+2a)}\Big]^2H(Q-a),\label{ref}\\
H(-a)&=&\Big[\frac{(b+6/b-2a)(3b/2+4/b-2a)}{(b/2+2a)(2/b+2a)}\Big]^2H(-Q'+a).\nonumber
\eeqa
Notice that these equations are invariant with respect to the symmetry
 $b\rightarrow-\frac{2}{b}$ with $a\rightarrow -a$ in agreement with the
well-known self-duality of the BD-model.

As was shown in the previous section, the solution of these functional
equations should exhibits, at least, the poles (\ref{jpole}) through the
identification $b=i\beta$ and $a=i\alpha$. Since the solution 
of (\ref{ref}) is defined up to a multiplication constant, we naturally 
choose to fix it by imposing eqs. (\ref{residu}). We find that the
``minimal'' solution which follows from these constraints is :
\beqa
H(a)&=&-\Big[\frac{m\Gamma(\frac{b^2}{h})\Gamma(\frac{2}{h})}{\Gamma(\frac{1}{3}){\sqrt 3}\ 2^{2/3+3/2}(Q+Q')^2}
\Big]^4\times
\frac{\gamma^2(\frac{1}{3})}{\gamma(\frac{2b^2}{h})\gamma(\frac{4}{h})}\label{VEVHreal}\\ \nonumber\ \ \
&\times& \ \
\gamma\big(\frac{2ba+4}{h}\big)\gamma\big(\frac{-2ba-b^2}{h}\big)
\gamma\big(\frac{2ba+3+b^2}{h}\big)\gamma\big(\frac{-2ba-1}{h}\big)\\ \nonumber \ \ \ 
&\times& \ \ 
\gamma\big(\frac{-2ba+2b^2}{h}\big)\gamma\big(\frac{2ba-2}{h}\big)
\gamma\big(\frac{-2ba+2+3b^2/2}{h}\big)\gamma\big(\frac{2ba-b^2/2}{h}\big)
\eeqa
where $h=6+3b^2$ is the ``deformed'' Coxeter number \cite{DGZ,CDS}. Here we
have used the exact relation between the parameters $\mu$ and $\mu'$ in
the action (\ref{actionBD}) and the mass of the particle $m$ \cite{5} : 
\beqa
m=\frac{2\sqrt 3 \Gamma(1/3)}{\Gamma(1+b^2/h)\Gamma(2/h)}
\big( -\mu\pi\gamma(1+b^2)\big)^{1/h} \big(
-2\mu'\pi\gamma(1+b^2/4)\big)^{2/h}.\label{massmu}
\eeqa
It is
easy to see that for $b=i\beta$ and $a=i\val$, ${\cal H}(\alpha)$  possess 
poles located at :
\beqa
\val_0\ + q(\frac{3}{\beta}-\frac{3\beta}{2})\ ,\ \ q\in{\mathbb Z} \ \
\ \mbox{with} \ \ \val_0\in\{-\frac{\beta}{2},\frac{1}{2\beta},\frac{\beta}{4},
\beta,\frac{3}{2\beta}-\frac{\beta}{2},-\frac{1}{\beta},
-\frac{1}{\beta}+\frac{3\beta}{4},\frac{2}{\beta}\}.\nonumber
\eeqa
But as long as we consider $\alpha$ that satisfy (\ref{cond}) there 
remain\,\footnote{Notice that for $\frac{2}{3}<\beta^2<1$ one also has the pole
$\alpha_0'\equiv -\frac{1}{\beta}+\frac{3\beta}{4}$. However, this one 
is nothing but obtained from the reflection $\alpha_0'=(\beta-\frac{1}{\beta})-\frac{\beta}{4}$.}
 the expected poles (\ref{jpole}) :
\beqa
\alpha_0\in\{-\frac{\beta}{2},\frac{1}{2\beta},\frac{\beta}{4},
\beta\}.
\eeqa

It is well-known that the Bullough-Dodd model at the specific value of the
coupling $b^2=2$ and the sinh-Gordon model at $b^2=\frac{1}{2}$
give an equivalent Lagrangian representation of the same QFT.
Then, as expected, one can check that (\ref{VEVHreal}) evaluated at $b^2=2$
coincides exactly with the same quantity in the ShG model evaluated at
$b^2=\frac{1}{2}$. 

Accepting the conjecture (\ref{VEVHreal}) and using eq. (\ref{L2L2}) for $a=0$ one
can easily deduce for instance : 
\beqa
<T{\overline T}>_{BD}\ \equiv\ <L_{-2}{\overline L}_{-2}{\mathbb I}>_{BD}\
=\ -\pi^2f^2_{BD}\label{TTbar}
\eeqa
where 
\beqa
f_{BD}&=&\frac{m^2}{16{\sqrt 3}\sin(\frac{\pi b^2}{h})\sin(\frac{2\pi}{h})}     
\eeqa
is the bulk free energy of the BD model \cite{5}. 

\resection{ Comparison with the semi-classical results}

As we saw previously, the OPE proposed in eq.
(\ref{twop}) plays a crucial role in the
determination of the prefactor of the $\alpha-$dependent part of ${\cal H}(\alpha)$, using eqs.
(\ref{residu}). It is therefore important to check this expression, using 
for instance the semi-classical expansion. In what follows, we will compare
(\ref{twop}) with the semi-classical calculations based on the action (1.1).

Let us consider (\ref{twop}) for $\alpha_1=\sigma/\beta$,
$\alpha_2=\omega\beta$ in the classical limit $\beta\rightarrow 0$. 
Then the saddle-point evaluation of the functional integral based on the action
(1.1) leads to the field configuration
$\varphi_{cl}=\frac{2}{i\beta}(\phi(t)+\frac{1}{3}\log(\frac{\mu'}{2\mu}))$, $t=\frac{mr}{2\sqrt 3}$, where
$\phi(t)$ is a solution of the classical Bullough-Dodd equation :
\beqa
\partial_t^2\phi +t^{-1}\partial_t\phi= 4(e^{2\phi}-
e^{-\phi})\label{BDcl}
\eeqa
with the following asymptotic conditions :
\beqa
\phi(t)&\rightarrow&-A\log t-\log B\label{tto0} + o(1) \ \  \ \qquad
\qquad \ \ 
\qquad \mbox{as}\ \ \ t\rightarrow 0\ ,\\
\mbox{and}\ \ \ \ \phi(t)&\rightarrow&
 -\frac{4{\sqrt
3}}{\pi}\sin(\frac{\pi\sigma}{3})\sin(\frac{\pi(1+\sigma)}{3})K_0(2{\sqrt
3} t) \ \ \ \mbox{as}\ \  \ t\rightarrow \infty\ .\nonumber
\eeqa
Here we denoted :
\beqa
A\equiv -2\sigma\ ; \ \ \ \ \
\ B\equiv \frac{3^{2\sigma}}{\gamma(\frac{1-\sigma}{3}) 
\gamma(\frac{2-2\sigma}{3})}\ ,\label{notation}
\eeqa
and $K_0(x)$ is the MacDonald function. Such a
solution was considered in \cite{TW}.

Taking into account the above considerations, the two-point function takes the
following form in the semi-classical limit :
\beqa
\frac{<e^{i\omega\beta\varphi}(x)e^{i{\sigma\over\beta}\varphi}(y)>_{BD}}
{<e^{i\omega\beta\varphi}>_{BD}<e^{i{\sigma\over\beta}\varphi}>_{BD}}\big|_{\beta^2\rightarrow 0} =
(e^{-2\phi})^{-\omega}.\label{twocl}
\eeqa
Following \cite{TW} it is not difficult to obtain the first few terms in the
$t\rightarrow 0$ expansion :
\beqa
e^{-2\phi(t)}\!\!\!\!&=&\!\!
B^2t^{2A}\Big( 1-{2\over B^2(A-1)^2}t^{2-2A} + {8B\over (A+2)^2}t^{2+A} + 
{40 B^2\over (A+2)^4}t^{4+2A} +\nonumber \\
             &+&\!\!\! {8(5A^2-4A-28)\over (A-1)^2(A+2)^2(A-4)^2B}t^{4-A} +
{1\over (A-1)^4 B^4}t^{4-4A} + O(t^{6+3A}\!,\! t^6)\Big).\label{expphi}
\eeqa

We would like now to compare these results with the corresponding limit of
(\ref{twop}). First, using
the result for the exact VEV in the Bullough-Dodd model (\ref{VEVBD}) proposed in \cite{5} we
obtain the following ratios :
\beqa
\frac{{\cal G}_{\sigma/\beta+\omega\beta}}{{\cal G}_{\sigma/\beta}{\cal
G}_{\omega\beta}}&=& m^{4\omega\sigma}
\frac{\Big(\gamma(\frac{1-\sigma}{3}) 
\gamma(\frac{2-2\sigma}{3})\Big)^{2\omega}}{2^{4\omega\sigma}3^{6\omega\sigma}};\nonumber
 \\
\mu^n\frac{{\cal G}_{\sigma/\beta+\omega\beta+n\beta}}{{\cal G}_{\sigma/\beta}{\cal
G}_{\omega\beta}}&=& \frac{m^{4\omega\sigma+4n\sigma+2n}}{\beta^{2n}\pi^n}
\frac{\Big(\gamma(\frac{1-\sigma}{3}) 
\gamma(\frac{2-2\sigma}{3})\Big)^{2\omega+2n}}
{2^{4\omega\sigma+4n\sigma+2n}3^{6\omega\sigma+6n\sigma+n}}; \\
{\mu'}^n\frac{{\cal G}_{\sigma/\beta+\omega\beta-n\beta/2}}{{\cal G}_{\sigma/\beta}{\cal
G}_{\omega\beta}}&=& \frac{m^{4\omega\sigma-2n\sigma+2n}}{\beta^{2n}\pi^n}
\frac{\Big(\gamma(\frac{1-\sigma}{3}) 
\gamma(\frac{2-2\sigma}{3})\Big)^{2\omega-n}}
{2^{4\omega\sigma-2n\sigma+n}3^{6\omega\sigma-3n\sigma+n}};\nonumber \\
\mu^n{\mu'}\frac{{\cal G}_{\sigma/\beta+\omega\beta+(n-1/2)\beta}}{{\cal G}_{\sigma/\beta}{\cal
G}_{\omega\beta}}&=& \frac{m^{4\omega\sigma+4n\sigma-2\sigma+2n+2}}{\beta^{2n+2}\pi^{n+1}}
\frac{\Big(\gamma(\frac{1-\sigma}{3}) 
\gamma(\frac{2-2\sigma}{3})\Big)^{2\omega+2n-1}}
{2^{4\omega\sigma+(4n-2)\sigma+2n+1}3^{6\omega\sigma+(6n-3)\sigma+n+1}}.\nonumber
\eeqa
Furthermore, the mass-$\mu$ relation (\ref{massmu}) proposed in \cite{5} gives :
\beqa
\mu{\mu'}^2\tend{\beta^2 \to 0} \Big(\frac{m}{2{\sqrt 3}}\Big)^6
\frac{4}{\pi^3\beta^6}
\eeqa
whereas, using eq. (\ref{jn}) and ${\cal F}_{1,1}(\alpha_1\beta,\alpha_2\beta,\beta^2)$
which can be deduced from the results of \cite{mag}, we have :
\beqa
&&j_n(\sigma,\omega\beta^2,\beta^2) \tend{\beta^2 \to 0}\ \ \ 
 \frac{\pi^n\beta^{2n}}{n!\ (1+2\sigma)^{2n}}
\frac{\Gamma(2\omega+n)}{\Gamma(2\omega)}\ ;\nonumber\\
&&j_n(-\sigma/2,-\omega\beta^2/2,\beta^2/4) \tend{\beta^2 \to 0}\ \ \ 
 \frac{\pi^n\beta^{2n}}{n!\ 2^{2n}(1-\sigma)^{2n}}
\frac{\Gamma(-4\omega+n)}{\Gamma(-4\omega)} \ ; \\
&&{\cal F}_{1,1}(\sigma,\omega\beta^2,\beta^2) \tend{\beta^2 \to 0}\ \ \
-\frac{\pi^2\beta^4\omega}{(1+2\sigma)^2(1-\sigma)^2(2+\sigma)^2}
\big[\sigma^2(2\omega+7)+\sigma(8\omega+10)+8\omega+1\big]\ ; \nonumber \\
&&{\cal H}(\sigma/\beta+\omega\beta) \tend{\beta^2 \to 0}\ \ \ 
O(\beta^4)\ .\nonumber
\eeqa
If we now use the same notations as above (\ref{notation}),
the semi-classical limit of the expression (\ref{twop}) takes the
following form :
\beqa
\frac{{\cal G}_{\sigma/\beta,\omega\beta}}{{\cal G}_{\sigma/\beta}{\cal
G}_{\omega\beta}}
\tend{\beta^2 \to 0}\ && t^{-2A\omega}B^{-2\omega} \times
 \Big[ 1+\frac{2\omega}{B^2(1-A)^2}t^{2-2A} - \frac{8\omega
B}{(2+A)^2}t^{2+A} + \frac{\omega(2\omega+1)}{B^4(1-A)^4}t^{4-4A}\nonumber \\
&& +\frac{8\omega(4\omega-1)B^2}{(2+A)^4}t^{4+2A}
-\frac{8\omega\big(A^2(2\omega+7)-A(16\omega+20)+32\omega+4\big)}{B(1-A)^2(2+A)^2(4-A)^2}
t^{4-A}\nonumber \\
&&\ \ \qquad \qquad + \ \ O(t^6)\ \ \ \Big].
\eeqa
It is straightforward to check that this result agrees perfectly with
(\ref{twocl}) through (\ref{expphi}).

\resection{Expectation values of the descendent fields in $\Phi_{12}$,
$\Phi_{21}$ and $\Phi_{15}$ perturbed minimal models}
For imaginary value of the coupling $b=i\beta$, $\mu\rightarrow-\mu$ and
$\mu'\rightarrow -\mu'$ the action of the BD model (\ref{actionBD})
becomes complex.
Whereas it is not clear if it can be defined as a QFT, this model is
known to be integrable and its $S$-matrix was constructed in
\cite{smir}. It is known that this model possess a quantum group symmetry
 $U_q(A_2^{(2)})$ with deformation parameter
$q=e^{i\frac{\pi}{\beta^2}}$ \cite{smir,cost}. An important role is played by
one of its subalgebras $U_q(sl_2)\subset U_q(A_2^{(2)})$. Following \cite{smir}
(see also \cite{5}), we can restrict the Hilbert space of states of the
complex BD model  at special values of the coupling constant, more precisely 
when $q$ is a root of unity, i.e. for :
\beqa
\beta^2=\frac{p}{p'} \ \ \ \ \ \ \ \ \ \ \mbox{or} \ \ \ \ \ \ \ \ \ \beta^2=\frac{p'}{p} \
\ \ \ \ \ \
\mbox{with} \ \ \ 1<p<p'
\eeqa
relative prime integers, in which case the complex BD is identified with the
perturbed minimal models (\ref{action}) or (\ref{actiontilde}),
respectively. In the following, $\Phi_{lk}$ will denote a primary field
of the minimal model ${\cal M}_{p/p'}$.

In the first case, the exact relation between the parameters $\lambda$
in (\ref{action}) and the mass of the fundamental kink $M$ can be found
 in \cite{5} with the result :
\beqa
{\lambda}^2=\frac{\pi^{\frac{\xi+4}{\xi+1}}}{2^{\frac{\xi+5}{\xi+1}}}\frac{\Gamma^2(\frac{\xi}{4\xi+4})
\Gamma(\frac{1}{2}-\frac{1}{\xi+1})}{\Gamma^2(\frac{3\xi+4}{4\xi+4})
\Gamma(\frac{1}{2}+\frac{1}{\xi+1})}
\Big[ \frac{M\Gamma(\frac{2\xi+2}{3\xi+6})}
{{\sqrt 3}\Gamma(\frac{1}{3})\Gamma(\frac{\xi}{3\xi+6})}\Big]^{\frac{3\xi+6}{\xi+1}}.
\eeqa
Here we denote 
\beqa
\xi=\frac{p}{p'-p}\ .\label{xi}
\eeqa
For unitary minimal models $\xi>1$ which, for ${\mathfrak
I}m(\lambda)=0$, corresponds to the massive phase \cite{5}. Using
the particle-breather identification :
\beqa
m=2M\sin\big(\frac{\pi\xi}{3\xi+6}\big)\label{mass}
\eeqa
and eqs. (\ref{L2L2}), (\ref{VEVHreal}) for imaginary coupling
$b=i\beta$ and parameter $a=i\big(\frac{l-1}{2\beta}-\frac{k-1}{2}\beta\big)$, it is
straightforward to get the VEV in the model associated with the action
 (\ref{action}) :
\beqa
\frac{<0_s|L_{-2}{\overline L}_{-2}\Phi_{lk}|0_s>}{<0_s|\Phi_{lk}|0_s>}
&=& -\Big[\frac{{\sqrt
3}\pi(\xi+2)M\Gamma(1+\frac{2+2\xi}{3\xi+6})}
{\Gamma(\frac{1}{3})2^{2/3+1/2}\Gamma(\frac{\xi}{3\xi+6})} \Big]^4
\frac{\gamma^2(1/3)}{\gamma(-\frac{2\xi}{3\xi+6})
\gamma(\frac{4+4\xi}{3\xi+6})}\nonumber\\
&&\ \ \ \ \ \ \ \ \ \ \ \ \ \ \ \ \ \ \ \ \ \ \ \ \ \ 
\times\ \ \ {\cal W}_{12}((\xi+1)l-\xi k)\label{12}
\eeqa
with 
\beqa
{\cal W}_{12}(\eta)=\frac{1}{\xi^{2}(\xi+1)^{2}}
\times w(\eta;\ 5+4\xi,\ 4+2\xi,\ -1-2\xi,\ 1+\xi/2;\ 3\xi+6)\nonumber
\eeqa
where we introduce the useful notation :
\beqa
w(\eta;a_1,a_2,a_3,a_4;g)=\prod_{i=1}^{4}
\gamma\big(\frac{a_i+\eta}{g}\big)\gamma\big(\frac{a_i-\eta}{g}\big).\nonumber
\eeqa
Here $|0_s>$ is one of the degenerate ground
states of the QFT (\ref{action}) (see \cite{5} for a detailed discussion of
the vacuum structure of the model). Taking $\Phi_{lk}$ in (\ref{12}) to be the
identity operator, it is easy to get :
\beqa
<T{\overline T}>=-\frac{\pi^2M^4}{48}\frac{\sin^2(\frac{\pi\xi}{3\xi+6})}
{\sin^2(\frac{\pi(2\xi+2)}{3\xi+6})}\ .\label{tt12}
\eeqa
A simple check consists to consider the scaling Lee-Yang
model which corresponds to $p=2$, $p'=5$ i.e. $\xi=\frac{2}{3}$ in (\ref{action}). As
$\Phi_{12}\equiv\Phi_{13}$ for these values, we must obtain the result
of
 \cite{des}. Using (\ref{mass}) the lightest mass in (\ref{action}) is :
\beqa
m=2M\sin(\frac{\pi}{12})\nonumber
\eeqa
Replacing this expression in (\ref{12}) for $l=1$, $k=3$ and (\ref{tt12}) it is easy to see
that the results are in perfect agreement with the ones of \cite{des}.

In the second restriction $\beta^2=p'/p$, which leads to the
action (\ref{actiontilde}). The exact relation between the parameter $\hat\lambda$
and the mass of the fundamental kink $M$ is in this case
\cite{5} :
\beqa
{\hat\lambda}^2=\frac{\pi^{\frac{\xi-3}{\xi}}}{2^{\frac{4-\xi}{\xi}}}
\frac{\Gamma^2(\frac{\xi+1}{4\xi})
\Gamma(\frac{1}{2}+\frac{1}{\xi})}{\Gamma^2(\frac{3\xi-1}{4\xi})
\Gamma(\frac{1}{2}-\frac{1}{\xi})}
\Big[ \frac{M\Gamma(\frac{2\xi}{3\xi-3})}
{{\sqrt 3}\Gamma(\frac{1}{3})\Gamma(\frac{\xi+1}{3\xi-3})}\Big]^{\frac{3\xi-3}{\xi}}.
\eeqa
Along the same line as for the $\Phi_{12}$ perturbation we obtain the
following expression for the VEV in
the model associated  with the action (\ref{actiontilde}) :
\beqa
\frac{<0_s|L_{-2}{\overline L}_{-2}\Phi_{lk}|0_s>}{<0_s|\Phi_{lk}|0_s>}
&=& -\Big[\frac{{\sqrt
3}\pi(1-\xi)M\Gamma(1-\frac{2\xi}{3-3\xi})}
{\Gamma(\frac{1}{3})2^{2/3+1/2}\Gamma(-\frac{\xi+1}{3-3\xi})} \Big]^4
\frac{\gamma^2(1/3)}{\gamma(\frac{2\xi+2}{3-3\xi})
\gamma(\frac{-4\xi}{3-3\xi})}\nonumber\\
&&\ \ \ \ \ \ \ \ \ \ \ \ \ \ \ \ \ \ \ \ \ \ \ \ \ \
 \times\ \ \ {\cal W}_{21}((\xi+1)l-\xi k)
\eeqa
with 
\beqa
{\cal W}_{21}(\eta)=\frac{1}{\xi^{2}(\xi+1)^{2}}
\times w(\eta;\ 1-4\xi,\ 2-2\xi,\ 1+2\xi,\ 1/2-\xi/2;\ 3-3\xi)\nonumber
\eeqa
where $|0_s>$ is one of the degenerate ground
states \cite{5} of the QFT (1.3). The analog of the formula (\ref{tt12}) is
now :
\beqa
<T{\overline T}>=-\frac{\pi^2M^4}{48}\frac{\sin^2(\frac{\pi(1+\xi)}{3-3\xi})}
{\sin^2(\frac{2\pi\xi}{3-3\xi})}\ .
\eeqa

Another subalgebra of $U_q(A^{(2)}_2)$ is the subalgebra
$U_{q^4}(sl_2)$.
One can again restrict the phase space of the complex BD with respect to this
subalgebra for a special value of the coupling :
\beqa
\beta^2=\frac{4p}{p'} \ \ \ \ \ \  \ \ \ \mbox{with} \ \ \ \ \ \ 2p<p'
\eeqa
relative prime integers. Then, for this value of the coupling, the BD
model is identified with the perturbed minimal model with the action
(\ref{actionhat}). The exact relation between $\tilde\lambda$ and the
mass\,\footnote{The general vacuum structure in the model
(\ref{actionhat}) is not clearly understood. However it is expected that
it possesses particles and kinks similarly to the other models (see for
instance refs. \cite{tak2} for details). The
physical mass scale $m$ is then associated with one of its particles \cite{5}.}
$m$ is \cite{5} :
\beqa
{\tilde\lambda}^2=\frac{(1+\xi)^2}{2\pi(1-4\xi)(1-2\xi)}
\frac{\Gamma(\frac{4\xi}{1+\xi})}{\Gamma(\frac{3-\xi}{1+\xi})}
{\sqrt {\frac{\Gamma(\frac{1}{1+\xi})\Gamma(\frac{5\xi}{1+\xi})}
{\Gamma(\frac{\xi}{1+\xi})\Gamma(\frac{1-4\xi}{1+\xi})}           }}
\Big[ \frac{m\Gamma(\frac{3-5\xi}{3-3\xi})\Gamma(\frac{1+\xi}{3-3\xi})}
{2{\sqrt 3}\Gamma(\frac{1}{3})}\Big]^{\frac{6(1-\xi)}{1+\xi}}.
\eeqa
Here we keep the definition (\ref{xi}). In particular, the massive
phase corresponds to :
\beqa
&&0\ <\ \xi\ <\ \frac{1}{4}\ ,\ \ \ \ {\mathfrak I}m {\tilde\lambda}=0\
;\nonumber \\
&&\frac{1}{4}\ <\ \xi\ <\ \frac{3}{5}\ ,\ \ \ \ {\mathfrak R}e {\tilde\lambda}=0\
.\nonumber 
\eeqa
As for the two previous cases one would like to obtain the expectation values
of the descendent fields for {\it any} primary operator $\Phi_{lk}$. For
$l>1$ these fields are not invariant with respect to the subalgebra
$U_{q^4}(sl_2)$ on the contrary to $\Phi_{1k}$. However, one expects that
they only differ by a c-number coefficient characterizing the degenerate structure of
the vacua\  $|0_s>$. Taking the ratio of the VEV of the descendent field of
$\Phi_{lk}$ 
associated with the action
(\ref{actionhat}) and the VEV of the primary field itself, one obtains :
\beqa
\frac{<0_s|L_{-2}{\overline L}_{-2}\Phi_{lk}|0_s>}{<0_s|\Phi_{lk}|0_s>}
&=& -\Big[\frac{m\xi \Gamma(1+\frac{1+\xi}{3-3\xi})\Gamma(-\frac{2\xi}{3-3\xi})}
{\Gamma(\frac{1}{3}){\sqrt 3}2^{2/3+1/2}} \Big]^4
\frac{\gamma^2(1/3)}{\gamma(-\frac{4\xi}{3-3\xi})
\gamma(\frac{2+2\xi}{3-3\xi})}\nonumber\\
&&\ \ \ \ \ \ \ \ \ \ \ \ \ \ \ \ \ \ \ \ \ \ \ \ \ \
 \times\ \ \ {\cal W}_{15}((\xi+1)l-\xi k)
\eeqa
with 
\beqa
{\cal W}_{15}(\eta)=\frac{1}{\xi^{2}(\xi+1)^{2}} 
\times w(\eta;\ \xi+5,\ 4-4\xi,\ -1-5\xi,\ 1-\xi;\ 6-6\xi)\nonumber\ .
\eeqa
In particular, we have :
\beqa
<T{\overline T}>=-\frac{\pi^2}{768}\frac{m^4}{\sin^2(\frac{2\pi\xi}{3-3\xi})
\sin^2(\frac{\pi(1+\xi)}{3-3\xi})}\ .
\eeqa

\resection{Concluding remarks}
In conclusion, we proposed in this paper an exact expression for the VEV of
the second level descendent of the exponential field 
$<(\partial\varphi)^2(\bar\partial\varphi)^2e^{i\alpha\varphi}>$ in the BD
model. The calculation is based on the so-called ``reflection relations'' which
lead to a system of functional equations for this VEV. While the solution is
not unique, we chose the ``minimal solution'' obeing some residue conditions. By
performing a quantum group restriction in the case of complex BD model
we found also the VEVs of the descendents of the primary fields in the
perturbed minimal CFT models
(1.2), (1.3), (1.4).

It is rather interesting to notice that in eq. (\ref{TTbar}), the exact VEV
$<T{\overline T}>_{BD}$ is simply related to the VEV of the trace of the energy momentum
tensor :
\beqa
\Theta=\frac{1}{4}T_{\nu}^{\nu}=\pi(1-\Delta_{pert})\mu\Phi_{pert}\nonumber
\eeqa 
where \ $1-\Delta_{pert}=h/4$\ \ and \
$\Phi_{pert}=e^{-\frac{b}{2}\vph}$\ \ as follows :
\beqa
<T{\overline T}>_{BD} = -<\Theta>^2_{BD}.
\eeqa
This was already noticed in \cite{des} for the ShG case. We expect this
property to be general, i.e. to be confirmed for other integrable
theories. However, we have no proof yet of this phenomena.

We would like to notice two important differences between ShG and BD models.
First, in the $\beta^2$ expansion of the two-point function the quantity 
${\cal H}{(\alpha)}$ (\ref{H}) comes with a coefficient of order $\beta^2$.
Therefore, it cannot be checked directly in the semi-classical approximation,
although the later is in agreement with the short distance expansion of the
two-point function, thus giving a strong support to our conjecture (\ref{VEVHreal}). For a direct
check one has to go beyond the classical limit and consider the first order in
the perturbation theory based on the action (1.1). 

Another difference is the
appearing of the third level descendents in the OPE of the exponential fields.
As a consequence, the following quantity appear in the short distance expansion
of the two-point function :
\beqa
{\cal K}(\alpha)=\frac{<(\partial\vph)^3({\overline\partial}\vph)^3e^{i\alpha\vph}>_{BD}}
{<e^{i\alpha\vph}>_{BD}}\label{K}.
\eeqa
In contrast with ${\cal H}(\alpha)$ it is sensitive to the semi-classical
expansion - it combines with the integral ${\cal F}_{1,2}$ in (\ref{twop}) in
order to match the corresponding term coming from the semiclassical calculation
based on the action (1.1). The function (\ref{K}) can also be obtained using
the ``reflection relations'' approach.

Using our results, one can deduce easily the next-to leading
contributions in the short distance behaviour of the two-point functions
 of primary operators for different perturbed minimal models : for
instance the Ising model in a magnetic field \cite{mag1} , the tricritical Ising
model perturbed by its leading energy density operator (with conformal
dimension $\Delta_{12}=1/10$) \cite{mag} or perturbed by its subleading magnetic
operator (with conformal dimension $\Delta_{21}=7/16$), and so on.

Several models can be worked out along the same line, using the known
results for the three-point functions of the CFT. For instance, the
super ShG model or, more generally, the parafermionic ShG model \cite{paraf1,9bis}.

We intend to discuss these various questions in a forthcoming publication
\cite{BS}.

\paragraph*{Aknowledgements}
We are grateful to Al. B. Zamolodchikov
 and particularly V.A. Fateev for valuable discussions and
interest in this work. We thanks for the hospitality of LPM (Montpellier)
 where part of this work was done. M.S. aknowledges the Physics
Department of Bologna University and APCTP, Seoul, for hospitality and
financial support. MS's work is supported under contract 
KOSEF grant 1999-2-112-001-5. PB's work is supported in part by the EU under contract 
ERBFMRX CT960012 and Marie Curie fellowship HPMF-CT-1999-00094.\\

\end{document}